# Parallel magnetic field suppresses dissipation in superconducting nanostrips

Yong-Lei Wang *,†,‡, Andreas Glatz *,§, Gregory J. Kimmel *,¶, Igor S. Aranson *,ǁ, Laxman R. Thoutam *,§, Zhi-Li Xiao *,§, Golibjon R. Berdiyorov **, François M. Peeters ††, George W. Crabtree *,‡‡, and Wai-Kwong Kwok *

*Materials Science Division, Argonne National Laboratory, Argonne, Illinois 60439, USA, †Department of Physics, University of Notre Dame, Notre Dame, Indiana 46556, USA, ‡School of Electronic Science and Engineering, Nanjing University, Nanjing 210093, China, §Department of Physics, Northern Illinois University, DeKalb, Illinois 60115, USA, ¶Department of Engineering Sciences and Applied Mathematics, Northwestern University, Evanston, Illinois 60208, USA, ǁDepartment of Biomedical Engineering, Pennsylvania State University, University Park, Pennsylvania, 16802, USA, **Qatar Environment and Energy Research Institute, Qatar Foundation, P.O. Box 5825, Doha, Qatar, ††Departement Fysica, Universiteit Antwerpen, Groenenborgerlaan 171, B-2020 Antwerpen, Belgium, and ‡‡Departments of Physics, Electrical and Mechanical Engineering, University of Illinois at Chicago, Illinois 60607, USA

The motion of Abrikosov vortices in type-II superconductors results in a finite resistance in the presence of an applied electric current. Elimination or reduction of the resistance via immobilization of vortices is the "holy grail" of superconductivity research. Common wisdom dictates that an increase in the magnetic field escalates the loss of energy since the number of vortices increases. Here we show that this is no longer true if the magnetic field and the current are applied parallel to each other. Our experimental studies on the resistive behavior of a superconducting $Mo_{0.79}Ge_{0.21}$ nanostrip reveal the emergence of a dissipative state with increasing magnetic field, followed by a pronounced resistance drop, signifying a re-entrance to the superconducting state. Large-scale simulations of the three-dimensional time-dependent Ginzburg-Landau model indicate that the intermediate resistive state is due to an unwinding of twisted vortices. When the magnetic field increases, this instability is suppressed due to a better accommodation of the vortex lattice to the pinning configuration. Our findings show that magnetic field and geometrical confinement can suppress the dissipation induced by vortex motion and thus radically improve the performance of superconducting materials.

superconducting vortices | parallel magnetic field | Time-dependent Ginzburg-Landau simulation

Abbreviations: TDGL simulation:Time-Dependent Ginzburg-Landau simulation

**V**ortex dynamics determines the electromagnetic responses of almost all practically important superconductors. Abrikosov vortices are created by the magnetic field penetrating a type-II superconductor, each carrying a single flux quantum surrounded by circulating supercurrents [1]. Understanding the electromagnetic properties of superconductors in applied magnetic fields is crucial for the majority of superconducting applications. If an electric current, **I**, is applied to the superconductor with a magnetic induction **B**, the associated Lorentz force $F_L = I \times B$ can induce motion of the vortices if it is greater than the vortex pinning force due to defects. [2] The vortex motion in turn leads to dissipation and breakdown of the zero-resistance state. Another practically important situation is when the magnetic field and the current are parallel, for example in force-free superconducting cables [3]. A tacit assumption is that in this case there will be no electric losses as vortices are typically aligned with the magnetic field, resulting in a zero Lorentz force configuration. However, previous experiments on superconductors in parallel magnetic fields are still controversial. Namely, an electric field, and, correspondingly, dissipation appears along the field/current direction [4–16]. Moreover, the superconducting critical current appears to increase with the magnetic field [4–8, 15]. Surprisingly, an instantaneous electric field *opposite* to the field/current direction was reported [9,11,12,16]. A number of theories were proposed to elucidate these phenomena, [13,17–32] including flux cutting [19,30,33,34], helical normal/superconducting domains, [35, 36] and helical vortex flow. [13, 31] However, the vortex behavior responsible for the observed dissipation in parallel magnetic field is still under debate [14, 28, 30, 31].

All the above mentioned experiments and theories focus on macroscopic samples with dimensions much larger than the superconducting coherence length. On the other hand, when the dimensions of the sample become comparable to the superconducting coherence length, the subtle interplay of vortex lattice confinement and pinning may lead to new behavior. Here, we investigate this non-trivial problem by carrying out experimental and theoretical investigations on superconducting strips with thicknesses comparable to the superconducting coherence length in parallel magnetic fields. We show that the resistance of the superconducting thin strips first increases with parallel magnetic field and then surprisingly drops back to zero at higher magnetic fields, indicating a re-entrance to the superconducting state. This novel resistive behavior exhibits strong dependences on temperature, current and strip thickness, which cannot be understood with any existing theory. Using large-scale, three dimensional (3D) time-dependent Ginzburg-Landau (TDGL) simulations [37, 38], we demonstrate that our observed resistive behavior is due to unwinding and reconnection of twisted vortices at intermediated parallel fields (see Fig. 1B). Further increase of the magnetic field results in straightening of vortices and increase in their constitutive pinning, leading to reduced dissipation.

## Sample configuration

The experiments were carried out on $Mo_{0.79}Ge_{0.21}$ (MoGe) superconducting thin strips with weak intrinsic vortex pinning [39]. Figure 1A shows the schematic of the sample orientation with respect to the magnetic field and the electrical current. The orientation of the magnetic field with respect to the sample was carefully aligned in a triple-axis vector magnet. The detailed field alignment procedures can be found in

[width=0.9]PNAS_Fig1.pdf

**Fig. 1.** Temperature and current dependent parallel magnetic field induced re-entrant superconductivity (A) Schematics of the superconducting MoGe film in a triple-axis vector magnet. The out-of-plane magnetic field $H_z$ is perpendicular to both the film plane and the current direction. The in-plane magnetic field $H_x$ is parallel to the film plane and perpendicular to the current direction. The in-plane magnetic field $H_y$ is parallel to both the film plane and the current direction. The electrical current is applied parallel to the $y$ direction. The inset shows a photo of the micro-patterned MoGe strip. The scale bar is 50 $\mu m$. (B) Magneto-resistance measured at $T = 5.8$ K for a 100 nm thick sample under magnetic fields in three orthogonal directions x (blue), y (red) and z (green), respectively. The resistance curve for parallel field ($H_y$) shows a resistant state at intermediate fields and a re-entrance of the superconducting state at higher fields. Experimental results (C,D) and TDGL simulations (E,F) for the parallel field dependent resistance at various temperatures (C,E) and at different currents (D,F). The zero field superconducting critical temperature and zero temperature upper critical field are $T_c = 6.2$ K and $H_{c2} = 6.7$ T, respectively. Other simulation parameters, impurity density and sample thickness, are $N_{imp} = 35$ and $L_z = \xi_0$ ($\xi_0 \approx 6 nm$ is zero temperature coherence length).

the Supplementary Information (SI). In this work, we mainly focus on the rarely investigated configuration of parallel magnetic field ($H_y$) and current in the sample plane. The bottom-left inset of Fig. 1A shows the micro-patterned MoGe strip.

For more details on the sample preparation, see Materials and Methods.

## Results and Discussion

**Parallel magnetic field induced re-entrant superconductivity.** The magneto-resistance of a sample with thickness of 100 nm is shown in Fig. 1B for magnetic fields applied in the three orthogonal directions with respect to the applied current and at a temperature of 5.8 K ($T_c \approx 6.2$ K). The green curve (with squares) corresponds to results for the out-of-plane magnetic field $H_z$ (perpendicular to the current). The resistance increases immediately with the magnetic field at very low fields due to the penetration of vortices, which are driven by the Lorentz force $\mathbf{F}_L$ and the weak intrinsic vortex pinning in MoGe [39]. When a magnetic field $H_x$ is applied in the sample plane (but perpendicular to the electrical current), the increase of the resistance with magnetic field is slower as compared to that for the out-of-plane magnetic field. Correspondingly, the zero-resistance state persists up to fields of 0.1 Tesla, as shown by the blue (with circles) curve in Fig. 1B. The zero resistance state is due to confinement effects induced by the surface barrier, whose effect is stronger in thinner samples [40]. Most previous investigations on superconducting vortex motion were conducted in the above two sample/field configurations while the Lorentz force-free configuration in parallel field was largely overlooked, especially in confined superconductors. The behavior corresponding to the magnetic field applied parallel to the applied electrical current ($y$-direction) is illustrated by the red (with triangles) curve in Fig. 1B. Here, the magneto-resistance first increases with $H_y$ at an even slower rate than that for $H_x$. With further increase of $H_y$, the resistance rapidly drops to zero (below the detectable resistance sensitivity of $10^{-5}$ $\Omega$), resulting in a re-entrance to the superconducting state.

Figures 1C and D show the temperature and current dependence of the observed re-entrance and panels E and F the corresponding simulation results. These will be discussed in detail below.

Cordoba et al. recently observed a reentrant dissipation-free state in W-based wires and TiN-perforated films [41]. In that work, the magnetic field and applied current are perpendicular to each other in the maximum Lorentz-force configuration and the reentrant dissipation-free state is due to arrested vortex motion by self-induced collective traps. This effect cannot explain our observed resistance behavior in parallel magnetic fields, where the Lorentz force is supposed to be absent. Furthermore, enhancement of superconductivity by a parallel magnetic field in an ultra-thin Pb-film and a two-dimensional electron gas at the interface of LaAlO$_3$/SrTiO$_3$ was reported in Ref. [42]. The enhancement was attributed to the *incre* of the mean-field critical temperature with increasing parallel magnetic field. Although enhanced superconductivity could lead to a gradual reduction of the dissipation, in our sample, the mean-field critical temperature *decreases* with magnetic field parallel to the current as demonstrated by the R-T curves in Fig. S4.

The emergence of a low dissipation state in our superconducting thin strips under parallel magnetic field is truly surprising and cannot be explained by any existing theories developed for macroscopic samples. For example, the helical vortex structures, as proposed for macroscopic superconducting wires or slabs [13, 31, 35, 36], cannot be formed due to the reduced sample thickness with highly confined vortices; Also, the result cannot be described by the flux cutting effect [19, 30, 33, 34], since it would require a very large current to produce a sufficient transverse magnetic field to tilt the vortices. Moreover, the current-induced magnetic field (self-field) can be safely neglected in our experiment at currents $I \leq 1$ mA.

**Suppression of vortex motion by parallel fields.** The resistance-temperature curves in semi-log plots show apparent kinks, as indicated by the arrows in the main panel of Fig. 2A and that of Fig. S4. The dissipation behavior below the kink, as compared to the rapid drop above the kink, can be attributed to vortex motion. At zero field, the dissipation is due to the motion of vortices and antivortices, which nucleate at the opposite outer boundaries and move towards the middle of the sample where they annihilate [43]. The temperature range for the transition below the kink at the intermediate field $H_y = 0.3$ T is wider than those measured at $H_y = 0$ T and 0.6 T. This indicates that the vortex motion is first enhanced with $H_y$ and then is significantly suppressed at higher fields. This is consistent with the resistance behavior shown in Fig. 1B for the parallel magnetic field $H_y$ dependence measured at a fixed temperature. The dissipation transition induced by vortex motion (below the kink) measured with a current of 1.0 mA (Fig. 2A) is more obvious as compared to that measured at a lower current (See Fig. S4B of corresponding measurement at 0.2 mA). This behavior is consistent with vortex motion driven by a Lorentz force, since higher currents produce stronger driving forces. Our observed dissipation state at intermediate $H_y$ most likely originates from vortex motion and the reentrant supercon-

[width=0.75]PNAS_Fig2.pdf

**Fig. 2.** Temperature dependence of vortex motion induced resistance behavior. (A) The temperature dependence of the resistance under parallel magnetic fields of 0 T (black-square), 0.3 T (red-dot) and 0.6 T (green-triangle) obtained with an applied current of 1.0 mA. The main panel in (A) is a semi-log plot and the inset shows the corresponding linear plot. The resistive states below the kinks pointed out by arrows are typical features originated from vortex motion. (B) Simulation results for the temperature dependence of the voltage (magneto-resistance) at different magnetic fields (0.08 - in the dissipationless, Lorentz-free regime, 0.13 - the intermediate resistive regime, 0.17 - in the re-entrant regime). The simulations are conducted using a current density of 0.18 and for a typical system size with $N_{imp} = 35$ inclusions.

ducting state could be ascribed to the suppression of vortex motion in a parallel field configuration. This "vortex crowding" effect means that a higher concentration of vortices can actually impede their motion.

Our measurements shown in Fig. 1B for the $H_{z\perp} I$ configuration indicate almost immediate breakdown of zero-resistance superconductivity, which would occur in relatively clean samples with small (but finite) concentration of point defects. In a parallel field, and in the absence of point disorder, vortex lines would be perfectly parallel to the applied field. Accordingly, the absence of a Lorentz force on the vortices should yield exceedingly high critical currents, which cannot lead to the observed intermediate dissipation state and re-entrant superconducting states in an ideally pure sample. However, in the presence of point disorder with finite concentration for a real sample, vortices can locally bend to accommodate the pinning sites, which results in the creation of a Lorentz force from the driving current. Consequently, vortex motion and resistive dissipation appear at a much smaller current density than that in the absence of disorder.

**Model and TDGL simulations.** To rationalize the experimental results and obtain insight into the behavior of superconducting currents in applied parallel magnetic fields, we performed large-scale vortex dynamics simulations based on the time-dependent Ginzburg-Landau equation (see Methods). In most of the simulations, the sample size was $128\xi_0$ in the field direction ($y$), with strip width ($x$ direction) of $64\xi_0$. The strip thickness is $16\xi_0 \approx 100$ nm, using zero temperature coherence length $\xi_0 \approx 6$ nm of MoGe estimated from the critical temperature versus magnetic field phase diagram [44]. The sample was periodic in the field direction ($y$). Superconductor/vacuum boundary conditions for the superconducting order parameter $\psi$, $\partial_z \psi = 0$, were imposed at the surfaces in the shorter transverse direction ($z$) in order to simulate the same thickness as in the experiment. The transverse size in the $x$ direction was chosen to be large enough such that the system behavior becomes insensitive to it, while maintaining a manageable computation time (see Methods).

Figure 1E shows our simulation results of the vortex dynamics for the parallel field dependent resistance at a fixed driving current ($j = 0.18$) and at different temperatures, while Figure 1F is simulated at fixed temperature ($T = 0.5T_c$) and at different driving currents. In both cases, the in-plane parallel field, $H_y$, is increased and the resulting dissipation, measured in the voltage drop across the sample, is shown. These results agree well with the experimental data shown in Fig. 1C and D.

In our model, we hypothesize that this phenomenon is related to a more optimal accommodation of the vortex lattice to disorder at low fields and straightening of the flux lines at high fields. A subsequent increase of the magnetic field results in a breakdown of the dissipation-free state and recovery of the resistive state due to vortex crowding, the mutual repulsion of vortices, and the geometric confinement by the sample boundaries results in straightening of the vortex configuration and mitigation of the bending due to defects. As indicated by Figs. 1E and 1F, the vortex dynamics behavior under parallel magnetic fields is strongly dependent on temperature and current. To further validate our hypothesis, we conducted experimental measurements and TDGL simulations at various temperatures and currents. The results are shown in Fig. 1C-1F. Remarkably, in faithful agreement between experiments and theoretical simulations, our results show an increase in temperature leads to narrowing of the re-entrant superconducting field-range and a slight decrease of the onset field. In other words, when the temperature rises, the resistance (or, equivalently voltage, which is proportional to the resistance) in the intermediate resistive state increases, while the field range of the re-entrant low dissipation state shrinks, as shown in Fig. 1C (experiment) and 1E (TDGL simulation). With further increasing temperature, the state becomes always resistive, although the voltage-magnetic field dependence is strongly non-monotonic (see brown-inverse triangle curves in Fig. 1C and in Fig. 1E). The effect of the applied current on the re-entrant superconductivity phenomena is shown in Figs. 1D (experiments) and 1F (TDGL simulation). When increasing the current, the intermediate resistance state in parallel magnetic fields is gradually enhanced while the magnetic field range of the re-entrant low-resistance state shrinks.

Fig. 2B shows the temperature dependence of the resistance/voltage at three different applied parallel fields: in the dissipationless, Lorentz-free regime(black), the intermediate resistive regime (red), in the re-entrant regime (green). Again, the behavior agrees very well with the observed behavior in the experiment (see Fig. 2A). We note that we chose the zero-temperature coherence length as the unit length, which results in a temperature dependent linear coefficient in the TDGL equation. As a reference point, we fixed this coefficient to be 1 (or $T = 0.5T_c$), resulting in a lower temperature than the experimental value. However, this allows us to change the temperature without altering the physical dimensions of the system, and at the same time, to obtain a reasonably large voltage signal. A direct reproduction of the experiments is impractical, since the linear coefficient in the temperature range of $T \sim 0.95 T_c$ is about 20x smaller and thus the voltage signal will be much noisier and more statistics would be required, making the simulations unfeasible.

For an intuitive understanding of the underlying processes resulting in the reentrance behavior, we studied the vortex dynamics at different fields. Snapshots highlighting the vortex configurations in the three distinctive regimes are shown in Figs. 3A-C. At relatively small magnetic fields, individual vortex lines are able to "optimize" deformations and accommodate to the given pinning sites (Fig. 3A) without the need to bend much. In this situation, the Lorentz force is not sufficient to overcome the pinning force and shows no dissipation. Increasing the magnetic field results in a larger number of deformed vortex lines (Fig. 3B). In the presented situation the vortex density becomes so large that vortices need to bend more at the pinning sites in order to be close to their equilibrium position (which would be an Abrikosov lattice without disorder). Due to this bending, the Lorentz force becomes larger than the pinning force and vortices start to move, i.e., the system becomes dissipative. The finite thickness of the sample initially allows the twisted vortices to unwind by bending towards the surfaces and move in opposite directions at the two opposing surfaces. If the field is increased even further (see Fig. 3C), the vortex density becomes even larger. In this case, the vortices do not have room to bend much and instead, they straighten due to vortex crowding. As a result, dissipation is suppressed and the voltage or magneto-resistance is decreased, constituting the reentrance region. When the magnetic field is above a certain value $H_{c*2}$, the critical current becomes lower than the applied current, leading to the emergence of the high-field resistive state.

The time dependence of the voltage in the intermediate resistive state is shown in Fig. 3D (see supplementary movie 3 for better visualization). In this resistive state, the overall dynamics is characterized by a periodic and relatively slow evolution of the vortex lines, resulting in a small voltage and fast reconnection/break up events manifested by sharp spikes

[width=0.7]PNAS$_F$ig3.pdf
**Fig. 3.** Time dependence of the vortex matter. Iso-surfaces of the order parameter magnitude $|\psi|^2$ illustrating snapshot vortex configuration in a rectangular sample for different magnetic fields $H_y$ in units of the zero-temperature upper critical field $H_{c2}(0)$. The green iso-surface for $|\psi|^2 = 0.4$ shows both tubes around vortex cores and blobs around pinning centers. Randomly-distributed point defects (seen as large green blobs) result in distorted vortex lines (seen as green tubes). (A) $H_y = 0.1$ in the pinned state where the vortex density is low and vortices are mostly parallel to the $y$-direction, (B) $H_y = 0.12$ in the periodic regime where vortices perform a helical motion along the surfaces, (C) $H_y = 0.18$ in the intermediate, low-resistance regime, where the vortex density is high such that vortex realignes with the $y$-axis and the resistance shows re-entrance behavior. Plot (D) shows the time evolution of the voltage (or magneto-resistance) in the periodic regime corresponding to panel (B). The time is measured in units of the Ginzburg-Landau time.

in the voltage time curve. However, the time scale of these oscillations is on the order of a few picoseconds (see Methods), such that a direct measurement is difficult. In the experiment, only the average voltage (DC measurement) is measured. All other plots of simulation data show voltage values averaged over long time intervals in steady state regimes.

We attribute this phenomenon to the subtle interplay between two competing factors that determine the vortex dynamics in superconducting nanostrips: geometrical confinement of vortices leading to an increase of the critical current and the distortion of vortex lines by point disorder resulting in a finite Lorentz force and subsequently to a decrease of the critical current. This interplay is strongly dependent on temperature and current as shown in Figs. 1C-F. In fact, increasing the temperature reduces the superconducting condensation energy and pinning forces induced by point defects, and thus enhances the intermediate dissipation states. Meanwhile, increasing the current also increases the driving force, which leads to enhancements of the intermediate dissipative states.

**Effect of strip thickness and defect concentration.** In our model, the vortex bending induced by finite point defects in a confined sample is the key for illustrating the parallel field induced dissipation state at intermediate fields and re-entrant superconducting state at higher fields. The vortex bending effect is expected to be sensitive to the sample thickness. To examine the effect of sample thickness, we fabricated MoGe strips with thicknesses ranging from 50 nm to 300 nm. Figure 4A shows the experimental results for samples with thicknesses of 100 nm (top panel), 200 nm (middle panel) and 300 nm (bottom panel). Increasing the thickness leads to a significant decrease of the re-entrant superconducting onset field. In other words, the dissipation states and re-entrant superconducting states emerge at lower fields for thicker samples. For the thinnest sample with thickness of 50 nm, we did not observe the re-entrant superconductivity effect at any of the tested temperatures or currents (see supplementary Fig.S3). The most likely explanation is that the re-entrance field of the 50 nm sample is higher than the upper critical field or beyond the highest experimentally accessible $H_y$. The effect of strip thickness can also be repeated by our TDGL simulation, as shown in Fig. 4B, which further confirms our hypothesis that the novel resistive state originates from motion of twisted/distorted vortices and the re-entrance of the superconducting state is induced by vortices straightening at higher parallel magnetic fields. The simulations confirm that the onset field $H_c{*}_1$ of the intermediate resistive region decreases with thickness, its width becomes smaller, and the maximum dissipation in this region decreases.

Since the appearance of the resistive state at lower fields is related to the local bending of vortices by point defects, it is clear that the re-entrance should depend on the concentration of defects in the sample (see Fig. 4C). Although it is difficult to investigate the effect of defect concentration experimentally, our TDGL simulation clearly indicates that there is an optimal defect concentration (about 35-40 impurities in the simulated system), where the critical field $H_c{*}_1$, when the sample first becomes resistive, is the smallest (see inset of Fig. 4C). Furthermore, at this optimal concentration, the re-entrance region is most pronounced with lowest intermediate resistance values and largest "second critical field", $H_c{*}_2$, (see Fig. 4C) above which the resistance starts to increase with the magnetic field again. This can be understood as follows: If the concentration is low, we are close to the Lorentz-force-free state with higher $H_c{*}_1$–$H_c{*}_2$ (see movie 1 in the SI). In contrast, if the concentration is too high, the effect of defects does not allow for an intermediate straightening of the vortices and decrease of the Lorentz force (see movies 2 in the SI) and therefore the system becomes resistive at $H_c{*}_1$ without a re-entrance region. Increasing the number of defects has the similar effect as increasing the temperature and narrows the re-entrant domain. These observations support our hypothesis that further increase in the number of vortices leads to straightening of vortices and decrease of the Lorentz force.

There are multiple roles of disorder (point defects) in superconductors. First, defects locally suppress superconductivity, which reduces the upper critical fields. Second, defects act as vortex pinning centers, which prevents vortex motion and increases the superconducting critical current [45]. Here, we investigate a third important consequence of defects in confined superconductors, namely the distortion or bending of vortex lines. The first two effects of defects do not change with the orientation of the magnetic field. However, the third effect, namely the local bending of vortex lines at defects plays a crucial role in a parallel magnetic field configuration as they lead to finite Lorentz forces, which can destroy the dissipationless state of a superconductor. This behavior is in stark contrast to a perpendicular field configuration, where point pinning impedes vortex motion and increases the critical current.

## Conclusion

We discovered a novel reentrant behavior of superconductivity in nanostrips placed in a parallel magnetic field. In experiments on MoGe samples, the magneto-resistance first increases with magnetic field, but at higher fields reduces to zero, such that superconductivity is recovered. This effect is strongly temperature dependent and can lead to a suppression of resistance below the measurable threshold over a range of a few kG. We elucidated the vortex dynamics and magneto-resistance behavior in the framework of large-scale, 3D time-dependent Ginzburg-Landau simulations. Our simulations revealed the mechanism for the observed behavior: the intermediate resistive state is due to a vortex instability leading to an unwinding of twisted vortex configurations. Upon increasing magnetic field, these instabilities are suppressed and the resistance drops due to a higher vortex concentration, leading to straightening of the vortex lines. An important factor in this interpretation is the presence of a small amount of defects in the system: without defects, vortices would simply align with the current until thermal fluctuations will bend them and the resulting Lorentz force will lead to a resistive state. This would happen at relatively high fields. The agreement

[width=0.9]PNAS_Fig4.pdf

**Fig. 4.** Effect of strip thickness and defect concentration. (A) Parallel field dependent resistance for samples with thickness of 100 nm (top panel), 200 nm (middle panel) and 300 nm (bottom panel). The corresponding sample and measurement parameters are listed in the figures. (B) TDGL simulation result of the dissipation voltage for two samples with different thicknesses: $16\xi_0$ and $24\xi_0$. The concentration of defects in both samples is similar indicated by the higher number of impurities in the thicker sample. (C) Simulation results for the parallel field dependence of the voltage (magneto-resistance) at fixed temperature $T/T_c = 0.5$ and applied current $j = 0.18$. The different curves show the dependence on the disorder (measured in the number of spherical impurities). Increase in the number of defects results in an overall increase of the voltage. The inset shows the non-monotonic dependence of the lower critical field onset, $H_{c1}^*$, of the resistive state as function of the number of impurities in the sample in a fixed size system.

between experiments and simulations on the evolution of the resistance as a function of magnetic field, temperature, electrical current and sample thickness indicates the close relevance of our simulations to the experimental data and validates our model.

## Materials and Methods

**Sample fabrication and measurement.** Films of 100 nm thickness were sputtered from a $Mo_{0.79}Ge_{0.21}$ alloy target onto silicon substrates with an oxide layer. The samples were patterned into 50 $\mu$m wide microbridges using photolithography. Transport measurements were carried out using a standard dc four-probe method with a Keithley 6221 current source and a Keithley 2182A nanovoltmeter. The electrical current flows horizontally along the long length of the microbridge. The applied magnetic field is precisely aligned along the electrical current direction in a triple-axis vector magnet. A detailed process of the field alignment can be found in the Supplementary Document.

**Simulation parameters.** For details on the algorithm and definition of all parameters, see Ref. [37]. The three-dimensional time-dependent Ginzburg-Landau equations

$$\partial_t \psi + i\mu\psi = E\psi - |\psi|^2\psi + (\nabla - iA)^2 \psi \quad [1]$$

$$j = \nabla\mu + \psi^*(\nabla\psi - iA), \nabla j = 0 \quad [2]$$

were solved numerically by an implicit finite-difference method. Here is the superconducting order parameter, $A = (0, 0, H_y x)$ is the vector potential corresponding to the applied magnetic field $H_y$ along y-direction, $j$ is the current density and $\mu$ is the scalar potential, $E = (T_c - T)/T$, $T$ is the temperature, $T_c$ is critical temperature. Note, that in this scaling the zero-temperature coherence length $\xi_0$ is the unit of length, the second critical field $H_{c2}(T=0)$ the unit for the field, and the depairing current density $j_{c0}$ for $H = 0$ is given by $j_{c0} = 2\sqrt{3}/9 \approx 0.385$. Here we used for the unit of current $j_0 = c\phi_0/(8\pi^2\xi\lambda^2)$ and the GL depairing current $j_{dp} = H_{c2}\xi/(6\pi\sqrt{3}\lambda^2)$.

The benchmark system has the dimensions: $L_x = 128\xi_0$, $L_y = 64\xi_0$, $L_z = 16\xi_0$ ($\xi_0 \approx 6$ nm) with open boundary conditions in z directions and periodic boundary conditions in x directions. The used computational mesh has $256 \times 128 \times 64$ grid points. The size and open boundary conditions in z direction are crucial for the confinement of the strip and for the study of the thickness dependence in reproducing the experimental system. The overall behavior is controlled by this smallest dimension of the system. For larger sizes (above $50\xi_0$), the behavior does not depend much on the thickness and the related boundary condition. Also, increasing the size or changing to periodic boundaries in y direction has no qualitative effect on the system's behavior. An external dimensionless current density of $j_x$ is applied in x-direction and the dimensionless magnetic field, $H_y$, is changed between 0 and 0.3 [in units of $H_{c2}(0)$]. The linear coefficient is set to $E = 1$ ($T/T_c = 0.5$).

Disorder is introduced by spherical defects with $E = -1.0$, corresponding to a normal region with $T_c = 0$, and diameter $4\xi_0$, due to its efficient pinning behavior [46]. $N_{imp}$ of those are randomly distributed in the simulation cuboid. The system is evolved over 25 million timesteps of length $0.1\tau_{GL}$, where $\tau_{GL} = \pi\hbar/(8k_B T_c)$ is the GL time - for MoGe this is given by $\tau_{GL} \approx 0.5$ps. After an initial relaxation period, the magnetic field is ramped from 0 to 0.3 in 100 steps. Before each ramping step, the voltage response is averaged over $10^5$ timesteps. The final magneto-voltage curves are then averaged over typical 12 disorder realizations.

The simulations were performed on Nvidia Tesla K20X GPUs. A magneto-voltage curve for a single disorder realization needs about 20 hours real simulation time.

**ACKNOWLEDGMENTS.** The computational work was supported by the Scientific Discovery through Advanced Computing (SciDAC) program funded by U.S. Department of Energy, Office of Science, Advanced Scientific Computing Research and Basic Energy Science, Division of Materials Science and Engineering. Experimental work was supported by the U.S. Department of Energy, Office of Science, Basic Energy Sciences, Materials Sciences and Engineering Division. L.R.T. and Z.L.X. acknowledge support through NSF Grant No. DMR-1407175. The nanopatterning and morphological analysis were performed at Argonne National Laboratory Center for Nanoscale Materials (CNM).

1. Abrikosov AA (1957) On the magnetic properties of superconductors of the second group. Sov. Phys. JETP 5:1174.
2. ANDERSON PW, KIM YB (1964) Hard superconductivity: Theory of the motion of abrikosov flux lines. Rev. Mod. Phys. 36:39.
3. Matsushita T, Kiuchi M, Otabe ES (2012) Innovative superconducting force-free cable concept. Superconductor Science and Technology 25(12):125009.
4. Sekula ST, Boom RW, Bergeron CJ (1963) Longitudinal critical currents in cold-drawn superconducting alloys. Applied Physics Letters 2(5):102.
5. Cullen GW, Cody GD, McEvoy JP (1963) Field and angular dependence of critical currents in $nb_3$ sn. Phys. Rev. 132:577.
6. Cullen GW, Novak RL (1964) Effect of fast neutron induced defects on the current carrying behavior of superconducting nb3sn. Applied Physics Letters 4(8):147.
7. LeBlanc M. A. R. , Belanger B. C., Fielding R. M. (1965) Paramagnetic helical current flow in type-ii superconductors. Phys. Rev. Lett. 14:704.
8. LeBlanc MAR (1966) Pattern of current flow in nonideal type-ii superconductors in longitudinal magnetic fields. Phys. Rev. 143:220.
9. Walmsley DG (1972) Force free magnetic fields in a type ii superconducting cylinder. Journal of Physics F: Metal Physics 2(3):510.
10. Nicholson J, Sikora P (1974) Flux flow in type ii superconducting wires in longitudinal magnetic fields. 17(3-4):275.
11. Walmsley DG, Timms WE (1977) Flux flow in longitudinal geometry. Journal of Physics F: Metal Physics 7(11):2373.
12. Cave JR, Evetts JE (1978) Static electric potential structures on the surface of a type ii superconductor in the flux flow state. Philosophical Magazine Part B 37(1):111.
13. Matsushita T, Ozaki S, Nishimori E, Yamafuji K (1985) A nonequilibrium themodynamic effect on a flux distribution in a superconductor under a longitudinal magnetic field. J. Phys. Soc. Jpn. 54(3):1060.
14. Clem JR, Weigand M, Durrell JH, Campbell AM (2011) Theory and experiment testing flux-line cutting physics. Superconductor Science and Technology 24(6):062002.
15. Irie F, Matsushita T, Otabe S, Matsuno T, Yamafuji K (1989) Critical current density of superconducting nbta tapes in a longitudinal magnetic field. Cryogenics 29(3, Supplement):317.
16. Matsushita T, Shimogawa A, Asano M (1998) Observation of structure of electric field on the surface of superconducting pbin slab under a longitudinal magnetic field. Physica C: Superconductivity 298:115.
17. Clem J (1975) On the breakdown of force-free configurations in type-ii superconductors. Physics Letters A 54(6):452.
18. Clem J (1977) Spiral-vortex expansion instability in type-ii superconductors. Phys. Rev. Lett. 38:1425.
19. Brandt E (1980) Continuous vortex cutting in type ii superconductors with longitudinal current. 39(1-2):41.
20. Clem J (1980) Steady-state flux-line cutting in type ii superconductors. 38(3-4):353.
21. Brandt EH (1982) Flux-line instability in a pin-free superconducting cylinder with longitudinal current. Phys. Rev. B 25:5756.
22. Clem JR, Perez-Gonzalez A (1986) Internal-magnetic-field distribution at the critical current of a type-ii superconductor subjected to a parallel magnetic field. Phys. Rev. B 33:1601.
23. Pérez-González A, Clem JR (1990) Flux-line-cutting effects at the critical current of cylindrical type-ii superconductors. Phys. Rev. B 42:4100.
24. Marsh GE (1994) Flux flow and flux cutting in type-ii superconductors carrying a longitudinal current. Phys. Rev. B 50:571.
25. Genenko YA (1995) Magnetic self-field entry into a current-carrying type-ii superconductor. ii. helical vortices in a longitudinal magnetic field. Phys. Rev. B 51:3686.
26. Aranson I, Gitterman M, Shapiro BY (1995) Onset of vortices in thin superconducting strips and wires. Phys. Rev. B 51:3092.
27. Shvartser M, Gitterman M, Shapiro BY (1996) Dynamics of helical vortices in a superconducting wire. Physica C: Superconductivity 264(3–4):204.
28. Ruiz HS, Badia-Mjaos A, Lopez C (2011) Material laws and related uncommon phenomena in the electromagnetic response of type-ii superconductors in longitudinal geometry. Superconductor Science and Technology 24(11):115005.
29. Ruiz HS, Lopez C, Badia-Majos A (2011) Inversion mechanism for the transport current in type-ii superconductors. Phys. Rev. B 83:014506.
30. Campbell AM (2011) Flux cutting in superconductors. Superconductor Science and Technology 24(9):091001.
31. Matsushita T (2012) Longitudinal magnetic field effect in superconductors. Japanese Journal of Applied Physics 51(1R):010111.
32. Genenko YA, Troche P, Hoffmann J, Freyhardt HC (1998) Chain model for the spiral instability of the force-free configuration in thin superconducting films. Phys. Rev. B 58:11638.
33. Campbell A, Evetts J (1972) Flux vortices and transport currents in type ii superconductors. Advances in Physics 21(90):199.
34. Blamire MG, Evetts JE (1986) Critical cutting force between flux vortices in a type-ii superconductor. Phys. Rev. B 33:5131.
35. Ezaki T, Irie F (1976) On the resistive state of current-carrying rods of type 2 superconductors in longitudinal magnetic fields. J. Phys. Soc. Jpn. 40(2):382.
36. Ezaki T, Yamafuji K, Irie F (1976) Flux flow in a current-carrying cylinder of type 2 superconductor in a longitudinal magnetic field. J. Phys. Soc. Jpn. 40(5):1271.
37. Sadovskyy I, Koshelev A, Phillips C, Karpeyev D, Glatz A (2015) Stable large-scale solver for ginzburglandau equations for superconductors. Journal of Computational Physics 294(0):639.
38. Kwok WK et al. (2016) Vortices in high-performance high-temperature superconductors. Rep. Prog. Phys. 79(11):116501.
39. Liang M, Kunchur MN, Hua J, Xiao Z (2010) Evaluating free flux flow in low-pinning molybdenum-germanium superconducting films. Phys. Rev. B 82:064502.
40. Bean CP, Livingston JD (1964) Surface barrier in type-ii superconductors. Phys. Rev. Lett. 12:14.
41. Crdoba R et al. (2013) Magnetic field-induced dissipation-free state in superconducting nanostructures. Nat Commun 4:1437.
42. Jeffrey Gardner H et al. (2011) Enhancement of superconductivity by a parallel magnetic field in two-dimensional superconductors. Nat Phys 7(11):895.
43. Latimer ML, Berdiyorov GR, Xiao ZL, Peeters FM, Kwok WK (2013) Realization of artificial ice systems for magnetic vortices in a superconducting moge thin film with patterned nanostructures. Phys. Rev. Lett. 111:067001.
44. Latimer ML, Berdiyorov GR, Xiao ZL, Kwok WK, Peeters FM (2012) Vortex interaction enhanced saturation number and caging effect in a superconducting film with a honeycomb array of nanoscale holes. Phys. Rev. B 85:012505.
45. Blatter G, Feigelman M, Geshkenbein V, Larkin A, Vinokur VM (1994) Vortices in high-temperature superconductors. Reviews of Modern Physics 66(4):1125.
46. Koshelev AE, Sadovskyy IA, Phillips CL, Glatz A (2016) Optimization of vortex pinning by nanoparticles using simulations of the time-dependent ginzburg-landau model. Phys. Rev. B 93:060508.

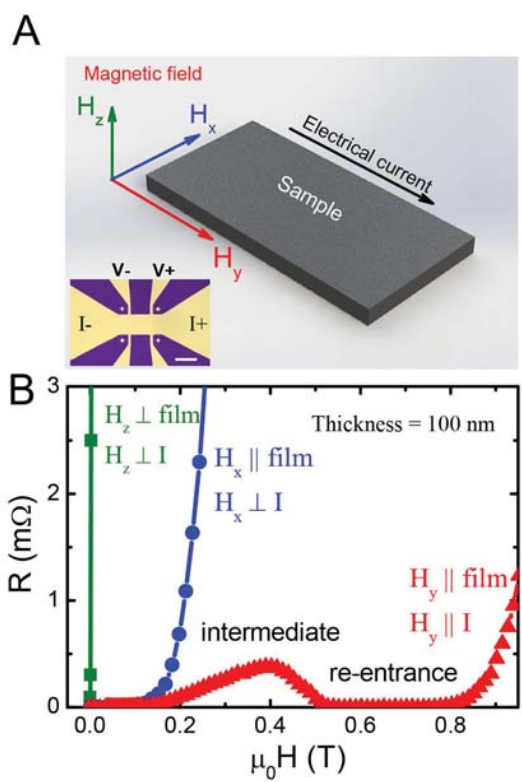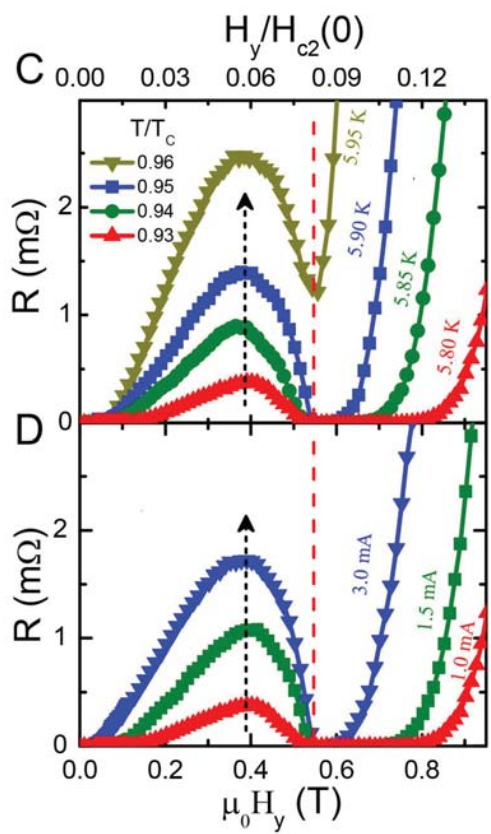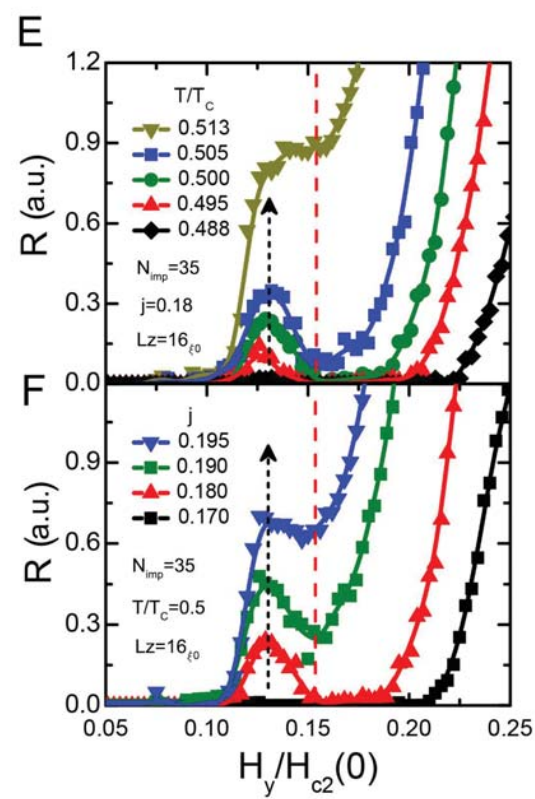

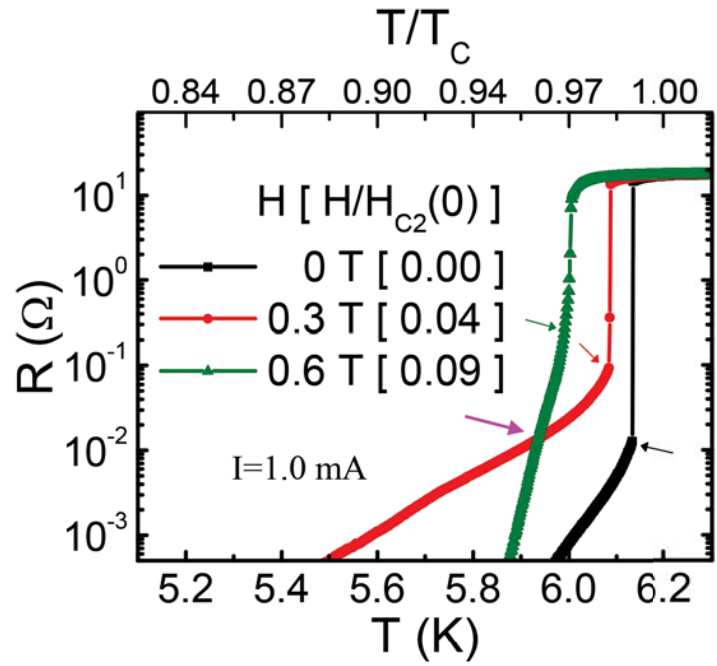

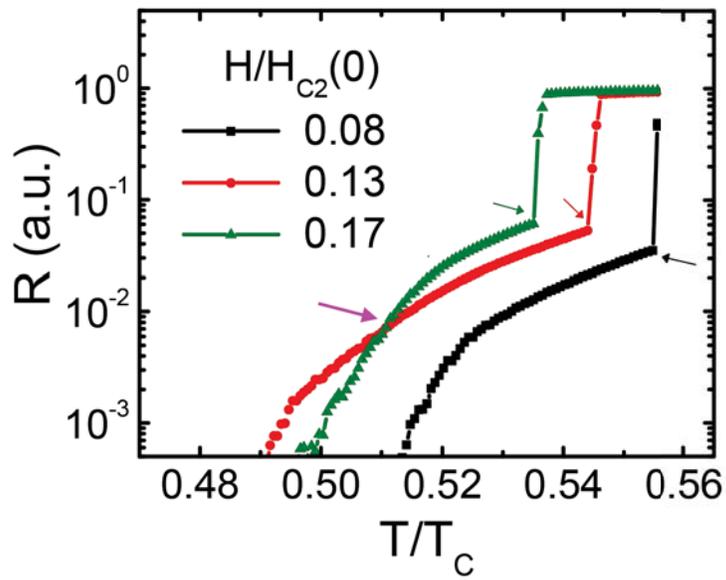

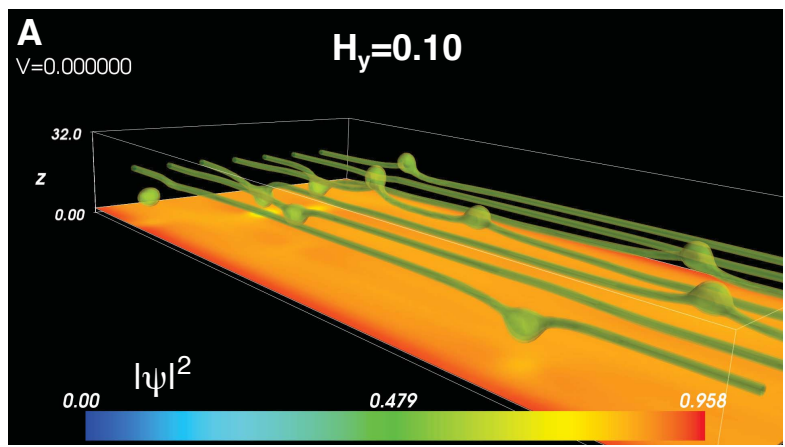
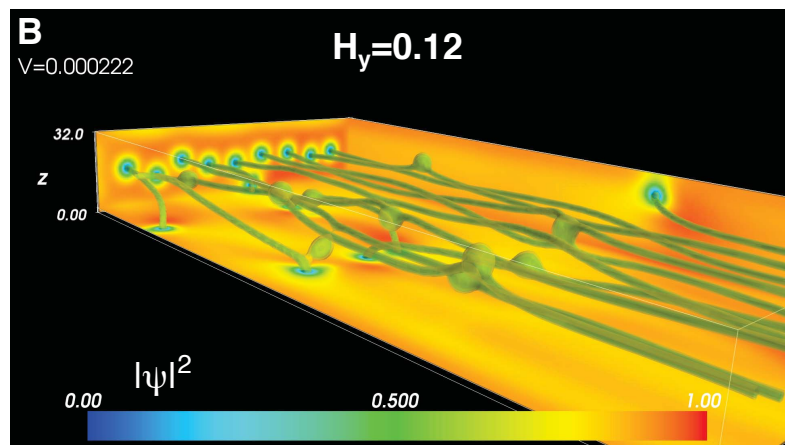
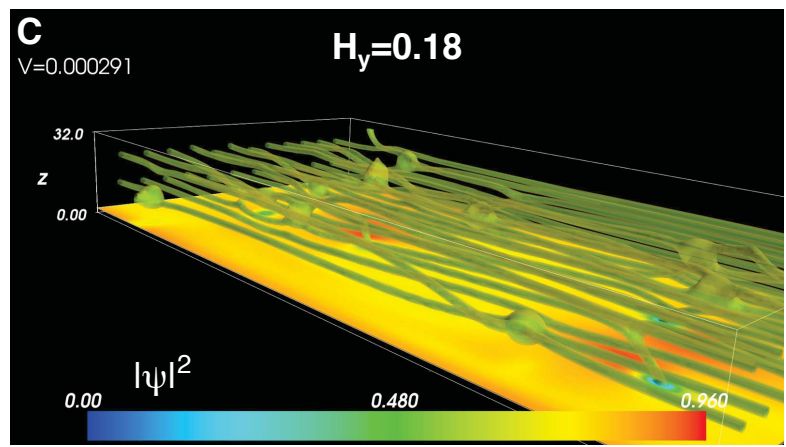
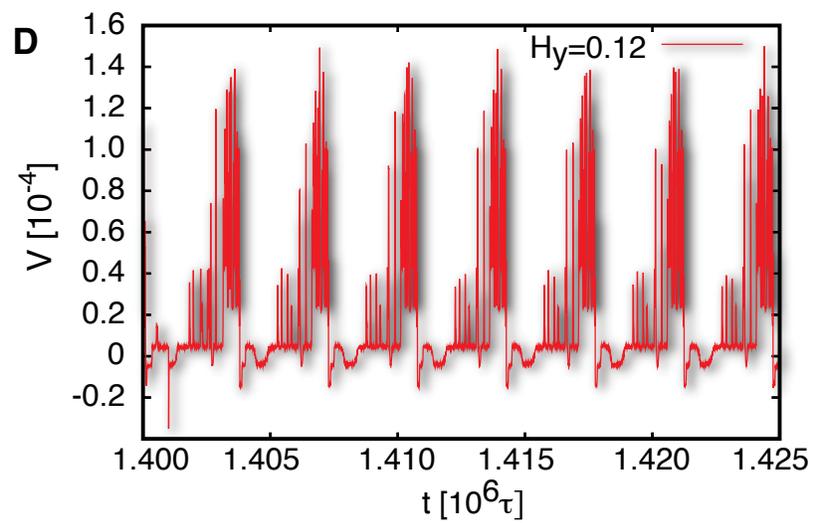

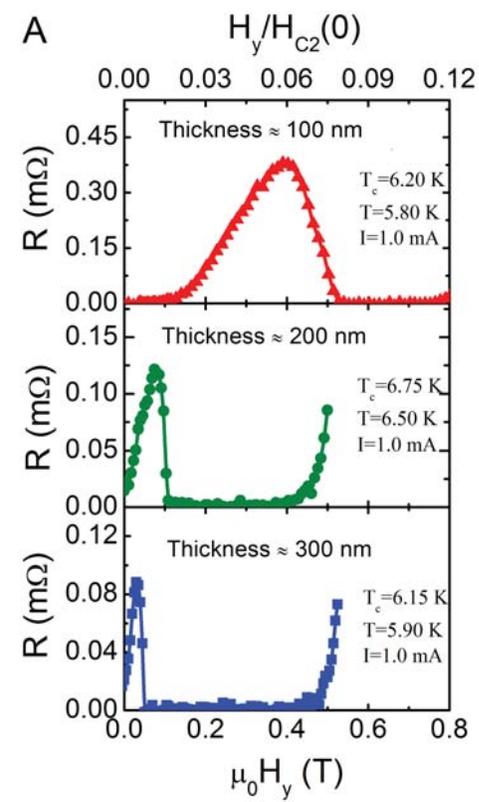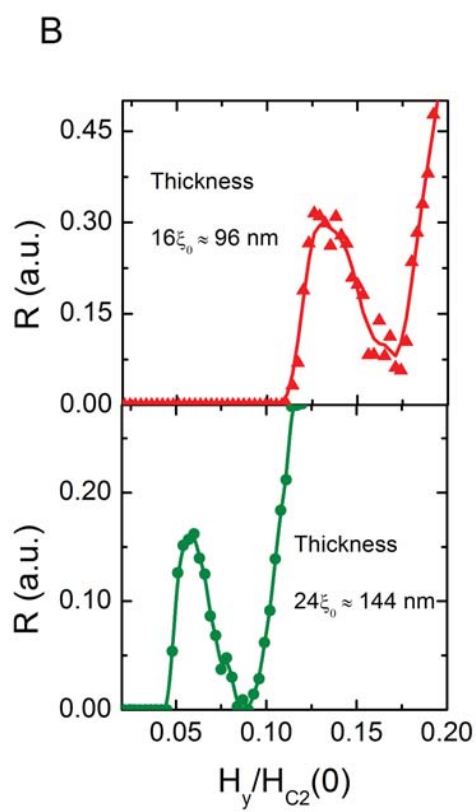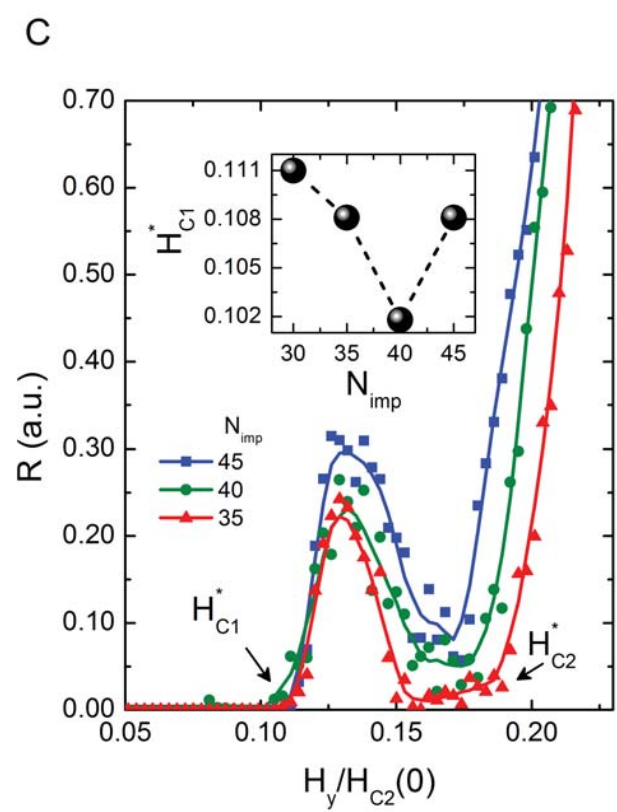